\documentclass[aps,prl,floatfix,twocolumn,footinbib]{revtex4}
\usepackage{epsfig}

\begin{document}

\title{Impact of Gas Giant Instabilities on Habitable Planets}
\author{Sonja Seppeur}
 
\affiliation{Institute for Theoretical Physics, Goethe
University, Frankfurt am Main, Germany}

\begin{abstract}
ABSTRACT\\
The detection of many extrasolar gas giants with high eccentricities indicates that dynamical instabilities in planetary systems are common. These instabilities can alter the orbits of gas giants as well as the orbits of terrestrial planets and therefore eject or move a habitable planet out of the habitable zone. In this work 423 simulations with 153 different hypothetical planetary systems with gas giants and terrestrial planets have been modelled to explore the orbital stability of habitable planets. The initial parameter variations include the number, distances and masses of the giant planets and the star type. Linear correlations between the initial number and the initial distances of gas giants and the survival rate of habitable planets were found. Planetary systems consisting of two giant planets are fairly benign to terrestrial planets, whereas six giant planets very often lead to a complete clearing of the habitable zone. Systems with initial distances of five Hill Radii between the giant planets have a high chance to harbour a habitable planet, although more compact systems are very destructive. The giant planet masses have a smaller impact on the stability of habitable worlds. Additionally, a link between the present-day orbit of an observed giant exoplanet and the survival rate of habitable planets was established. As a rule of thumb, observed gas giants with eccentricities higher than $0.4$ and inclinations higher than 20 degrees have experienced strong planet-planet scatterings and are unlikely to have a habitable planet in its system. Furthermore, it was found that habitable planets surrounding a K or M-star have a higher survival rate than those surrounding a G-star.

\end{abstract}\maketitle

\section{\label{sec:Intro} Introduction}

Since the first discovery of an extrasolar planet in 1992, the search and exploration of exoplanets is a fast-growing field of research. It is assumed that one important requirement for life is liquid water on the planetary surface, which is fulfilled in a certain distance range around a star called "the habitable zone". A rocky planet has to stay long enough in the habitable zone that life can evolve and survive. In a system with multiple planets these planets can potentially interact with each other. This can lead to periods of instability and chaos, which could be very destructive for terrestrial planets in the habitable zone. Until now, many extrasolar gas giants with high eccentricities have been discovered, which indicates that dynamical instabilities in planetary systems may be common. To explore the dynamical past of a planetary system, the system has to be simulated with specified initial conditions.\\

Many N-body simulations of planetary systems have been modelled yet. The so-called "Nice Model" is the most established model of our solar system. Tsiganis and Morbidelli (2005) modelled the past of the four giant planets of the solar system to explain among other things the today's eccentricities and inclinations of Jupiter, Saturn, Uranus and Neptune as well as many objects of the Kuiper belt. They found out that a compact, circular and co-planar configuration of the planets result in strong planet-planetesimal scattering and migration. After this phase of instability the outcome matches well the today's architecture of the outer solar system.\\

Similar simulations were made by Chambers and Kaib (2015). They additionally considered the evolution of terrestrial embryos and planets during the Nice Model starting with five and six giant planets. The results show that the possibility exists that one or two originally giant planets in our solar system could be ejected by strong gravitational scattering leaving the solar system in its current configuration. Several authors have simulated known extrasolar planetary systems. Chambers et al. (2000) investigated the stability of orbits of terrestrial planets in the habitable zones of the four known systems Rho CrB, 47UMa, Gliese 876 and Ups And. For this purpose one or two hypothetical terrestrial planets were placed in the habitable zone and the whole system was integrated for $10^8-10^9$ years to analyse the orbital stability.\\

A more extended study was made by Menou and Tabachnik (2002), who quantified the dynamical habitability of 85 known extrasolar planetary systems. They placed hundred test particles in the habitable zone and varied the masses of the other planets. They found out that one fourth of all systems retain a high percentage of their test particles. However, Matsumura et al. (2013) mentioned that it is also important to regard the dynamical history of a system. Some orbits that are stable today may have been unstable in the past and therefore may contain no terrestrial planet. They modelled three Jupiter mass planets and eleven test particles to determine unstable regions. The giant planets had orbit crossings after a few hundred years and were integrated for ten million years. The results show that a high percentage of test particles were removed due to secular perturbation of the giant planets, which implies that stable regions computed by the current distribution of giant planets are highly overestimated.\\

The most related work to this one is the publication of Carrera, Davies and Johansen (2016), who extended the work of Matsumura et al. (2013). They modelled systems with three Jupiters and systems with four giant planets with three different unequal mass sets and placed hundred test particles in the habitable zone. The three Jupiter systems were considerably more destructive to the test particles than the four unequal giant planets. Compared to Matsumura et al., they moved the giant planets outward and increased their separations, so that instability occurs after a few million years instead of a few hundred years. This timescale is more consistent with terrestrial planet formation. Carrera, Davies and Johansen also tested the effect of multiple terrestrial planets compared to test particles, which resulted in a decreasing survival rate of terrestrial planets due to dynamical interactions or physical collisions among themselves.\\

In this publication the work from Carrera, Davies and Johansen (2013) has been extended by considering plenty more initial conditions. Therefore, the survival rate of test particles in the habitable zone around different star types and with different initial numbers of giant planets between two and six planets are investigated, because it is not known how many planets form per a typical protostellar disk. In addition, the effect of different giant planet masses and distances are examined. The aim of this work is to get a probability estimate of the existence of habitable planets in exoplanetary systems with detected gas giants. To covering a broad range of possible giant planet architectures, many different initial conditions were simulated. Therefore a connection between the final parameters of the giant planets and the survival rate of test particles are produced. Another aim is the exploration of the correlation between the survival rate of test particles and the initial parameter variations. The results may be useful for future space missions to estimate the existence of a habitable planet in a system with already detected gas giants on the basis of their current orbital parameters.

\section{\label{sec:model} Methods}

N-body simulations of different hypothetical planetary systems were performed using the hybrid integrator of the MERCURY code (Chambers 1999).
Systems with gas giants in the outer region and test particles in the habitable zone were simulated. In the following the initial conditions are presented.\\

\paragraph{Test Particles}
In all runs twelve test particles are placed at equal distances in the habitable zone of the star to represent possible orbits of terrestrial planets. This is a good approximation because the mass of a minor planet is much smaller than the mass of a giant planet. However, test particles do not perturb each other or other bodies in the system, but several terrestrial planets would perturb each other and eventually destabilise their orbits. Therefore, test particles are an optimistic approximation for terrestrial planets in dynamical simulations.\\

\paragraph{Planet Number and Distances}

The giant planet number changes from two to six planets per planetary system. They are placed on initially coplanar, circular and compact orbits. Planet formation theory suggests that giant planets end up in a compact non-excited configuration after the dispersion of the gas disk.\\

The innermost giant is placed on three different positions dependent on the type of star. The other giant planets are placed at fixed separations in mutual Hill radii $\Delta$.\\

A planetary system is said to be Hill stable when the planet orbits can never cross. A system with two planets on circular, coplanar orbits is Hill stable if $\Delta > 2\sqrt{3}$, where $\Delta$ is the semimajor axis separation measured in mutual Hill radii.

\begin{equation}
\Delta =\frac{a_2-a_1}{R_H}
\end{equation}

\begin{equation}
R_H=\left[{\frac{m_1+m_2}{3M_{star}}}\right]^{\frac{1}{3}}\frac{a_1+a_2}{2}
\end{equation}

Systems with more than two planets cannot be Hill stable. For those systems with $N_{planet}>2$ no analytical criterion for absolute stability exists. The distances of the planets are calculated with equations 1 and 2 using $\Delta$, the semimajor axis of the neighbouring inside planet $a_1$ and the masses of the star and the two considered planets $M_{star},m_1,m_2$. For two planets $\Delta$ is chosen near the Hill stability boundary, because there is a sharp transition from Hill stability to instability (Table 1). For systems with more than two planets instability can occur at wider separations than in two-planet systems.\\

\renewcommand{\thetable}{\arabic{table}}

\begin{table}[h]
	\centering
	\begin{tabular}{c||l|l|l}
		$N_{planets}$ & 2 &  3,4,5,6 \\
		$\Delta$ & 3 3.2 3.4 & 4 4.5 5 \\
	\end{tabular}
	\caption{Defined mutual Hill separations for two, three, four, five and six planets. For two planets instability occurs at nearer separations than for systems with more than two planets.}
	\label{Delta}
\end{table}

\paragraph{Planet Masses}

Three different planet mass configurations for every number of planets were chosen, all with decreasing masses with increasing distance from the star. Carrera, Davies and Johansen simulated three Jupiter-mass planet systems and the results show that they were very destructive to terrestrial planets in the habitable zone compared to four giant planets with unequal masses. They also found out that the more hierarchical the masses of a system the higher is the survival rate of habitable planets. Therefore, only unequal masses were considered, because the planets in our solar system have unequal masses and this assumption matches better with planet formation theory. The first mass set has a low-mass ratio and the second one has a high mass-ratio starting with one Jupiter mass for the innermost planet. The third mass set has double mass of the high mass-ratio set. Set one is less hierachical than set two (Table 2). \\

\begin{table}[h]
	\footnotesize
	\centering
	\begin{tabular}{c||l|l|l}
		$N_{}$ & $m_{low ratio} [m_J]$ & $m_{high ratio} [m_J]$ & $m_{double high} [m_J]$ \\
		2 & 1 0.8 & 1 0.3 & 2 0.6 \\
		3 & 1 0.8 0.5 & 1 0.3 0.1 & 2 0.6 0.2 \\
		4 & 1 0.8 0.5 0.3 & 1 0.5 0.2 0.05 & 2 1 0.4 0.1 \\
		5 & 1 0.8 0.6 0.4 0.2 & 1 0.6 0.3 0.1 0.05 & 2 1.2 0.6 0.2 0.1 \\
		6 & 1 0.8 0.6 0.4 0.2 0.1 & 1 0.7 0.4 0.2 0.1 0.05 & 2 1.4 0.8 0.4 0.2 0.1\\
	\end{tabular}
	\caption{Defined masses of the giant planets in Jupiter masses $m_J$ for each planet number N. In every column the first value is the mass of the innermost planet and the following values are for the next planets with larger distances from the star, respectively.}
	\label{Massenplanets}
\end{table}

\paragraph{Remaining Orbital Parameters of Giant Planets and Test Particles}

The initial orbits are circular, which corresponds to an eccentricity of zero. To get nearly coplanar orbits each orbit gets a random inclination between $0^\circ$ and $5^\circ$ (following the prescription of Johansen et al. 2012). The argument of pericenter $\omega$, the longitude of ascending node $\Omega$ and the mean anomaly get a random value between $0^\circ$ and $360^\circ$.\\

\paragraph{Star Types}

In addition, systems with different types of stars were simulated to analyse the dynamical effect of the star mass on the stability of orbits of terrestrial planets in the habitable zone. Three different types of stars G, K and M were considered (Table 3). One major reason for this choice is the long lifetime of these stars, which is long enough for life evolution. The distance of the habitable zone changes with the type of star. The boundaries of the habitable zones for G, K and M-stars were calculated using the online available "Habitable Zone Calculator" (Kopparapu et al. 2013). The twelve test particles were distributed uniformly between the boundaries of the HZ (Table 4).\\\\

\begin{table}[h]
	\centering
	\begin{tabular}{c||l|l|l}
		Star & G & K & M \\
		 M & $1M_{sun}$ & $0.8M_{sun}$ & $0.3M_{sun}$ \\
		 L & $L_{sun}$ & $0.15L_{sun}$ & $0.01L_{sun}$ \\
		 T & $5800^\circ C$ & $(3500-4850)^\circ C$ & $(2000-3350)^\circ C$ \\
		HZ & $(0.75-1.765)AU$ & $(0.31-0.783)AU$ & $(0.1-0.267)AU$ \\
	\end{tabular}
	\caption{Masses M, luminosities L, temperatures at the surface T and calculated habitable zone limits HZ for G, K and M-stars are shown.}
	\label{startypes}
\end{table}

\begin{table}[h]
	\centering
	\begin{tabular}{c||l|l|l|}
		& G & K & M \\
		$a_1 (TP)$ & 0.7 & 0.3 & 0.08 \\
		$a_2 (TP)$ & 0.8 & 0.35 & 0.1 \\
		$a_3 (TP)$ & 0.9 & 0.4 & 0.12 \\
		$a_4 (TP)$ & 1.0 & 0.45 & 0.14 \\
		$a_5 (TP)$ & 1.1 & 0.5 & 0.16 \\
		$a_6 (TP)$ & 1.2 & 0.55 & 0.18 \\
		$a_7 (TP)$ & 1.3 & 0.6 & 0.2 \\
		$a_8 (TP)$ & 1.4 & 0.65 & 0.22 \\
		$a_9 (TP)$ & 1.5 & 0.7 & 0.24 \\
		$a_{10} (TP)$ & 1.6 & 0.75 & 0.26 \\
		$a_{11} (TP)$ & 1.7 & 0.8 & 0.28 \\
		$a_{12} (TP)$ & 1.8 & 0.85 & 0.3 \\
		\\
		$a_1$ (first planet) & 3 4 5 &  2 3 4 & 1.5 2.5 3.5 \\ 
	\end{tabular}
	\caption{The semimajor axis of the twelve test particles within the habitable zone boundaries of each type of star are shown in astronomical units (AU). In the last row the three different values for the semimajor axis of the innermost giant planet are shown. The HZ boundaries were slightly broadened for an optimistic consideration.}
	\label{HZstars}
\end{table}

Finally, without consideration of the different star types these variations result in 135 planetary systems with different initial conditions. Additionally, each system is simulated three times for three million years to receive an ensemble of outcomes with similar initial conditions, because a dynamically unstable planetary system is highly chaotic. For statistical purpose the mean values of the final parameters are considered and illustrated. Due to the long computing time each system could only be simulated three times.

\section{\label{sec:model} Results}

Figure 1 shows that in planetary systems with two gas giants in many cases almost all test particles stay in the habitable zone during the simulation. Usually those runs with 12 surviving test particles experienced no instability phases. Typically the first  planet ejects the second less massive planet, whereby a part of the test particles survive.\\

\begin{figure}[!htbp]
	\centering
	\includegraphics[width=0.5\textwidth]{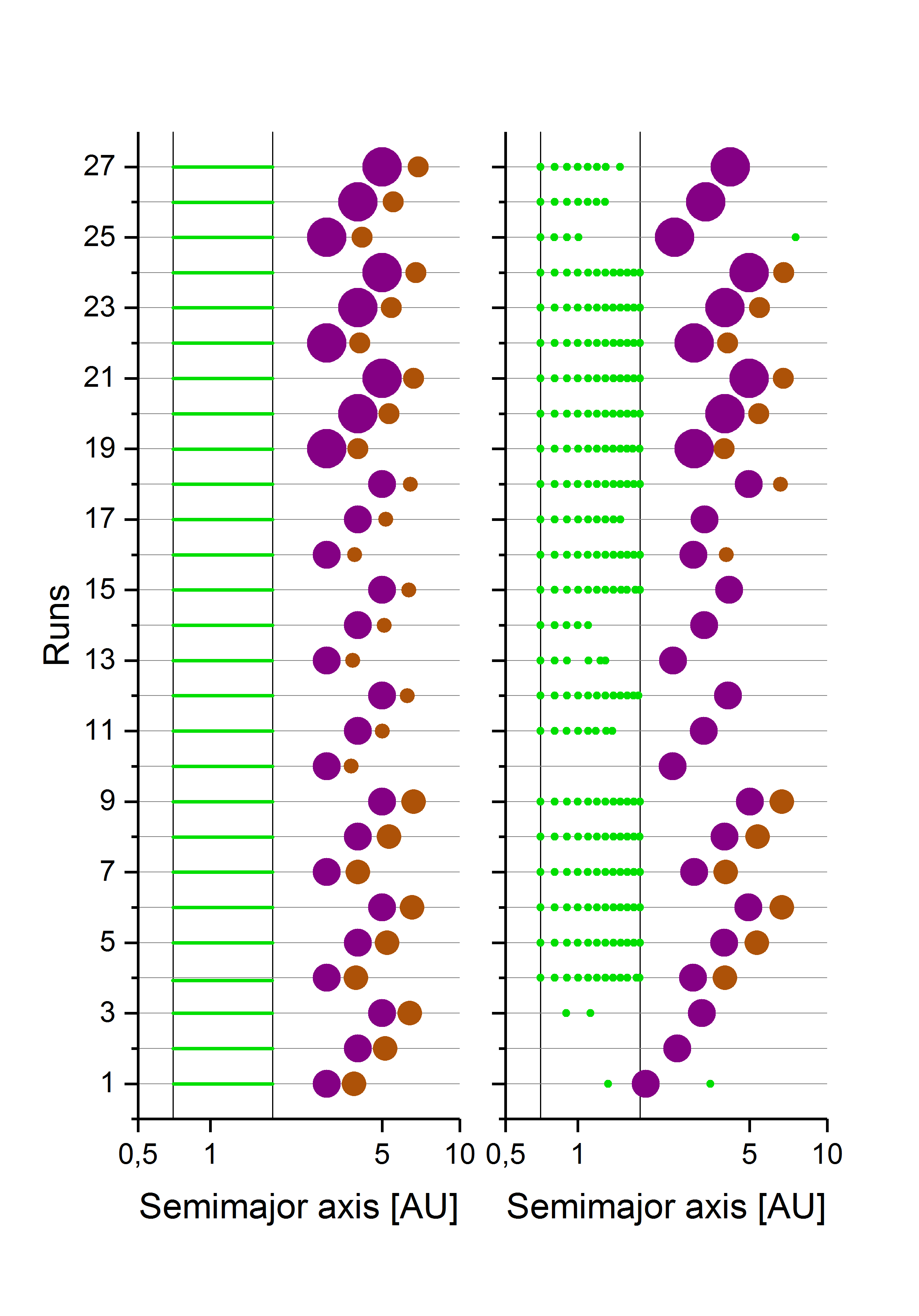}
	\caption{Illustration of the runs 1 to 27 with two planets. The left panel shows the initial state and the right panel shows the final state. The position in astronomical units is represented through a logarithmic x-axis and the size of the circles is proportional to the planet mass. The two vertical grey lines indicate the borders of the habitable zone ($0.7-1.8 AU$). The green lines (left plot) and accordingly the green dots (right plot) represent the twelve test particles.}
	\label{Results2pl}
\end{figure}

In about half of the runs in figure 2 the third planet got ejected. In the runs 28 to 31 the third planet got ejected and no test particle survived due to inward migration of the first planet very close to the habitable zone.\\

\begin{figure}[!htbp]
	\centering
	\includegraphics[width=0.5\textwidth]{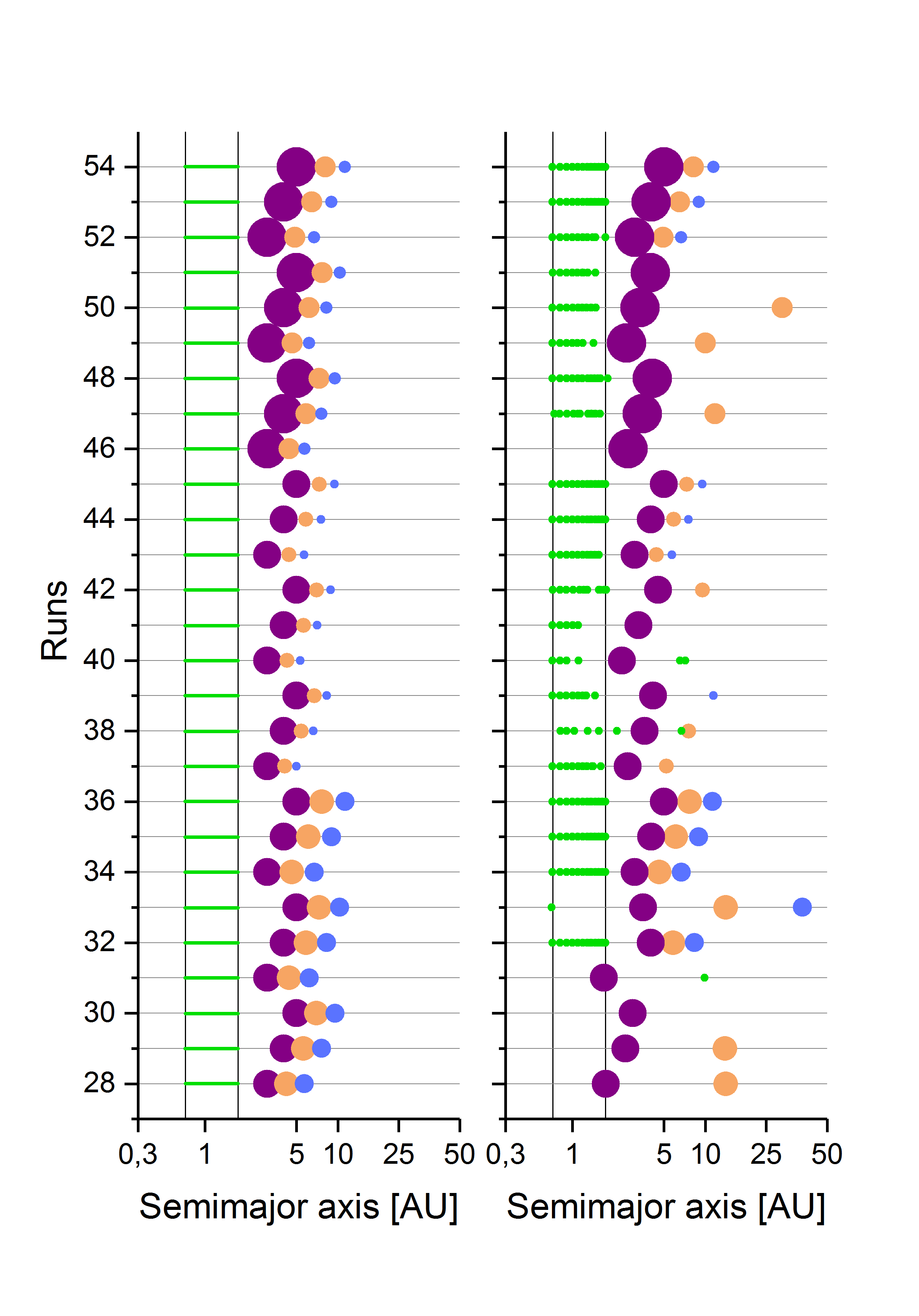}
	\caption{Same as figure 1 for three planets (runs 28-54).}
	\label{Results3pl}
\end{figure}

In figure 3 only two runs end up with the fourth planet and only seven runs with the third planet. In three runs the first and the second planet crossed their orbits and the second planet ended up nearer to the star than the first planet. This figure clearly shows the evacuation of the habitable zone when a giant moves inward close to it. Only run 80 and 81 experienced no instabilities.\\

\begin{figure}[!htbp]
	\centering
	\includegraphics[width=0.5\textwidth]{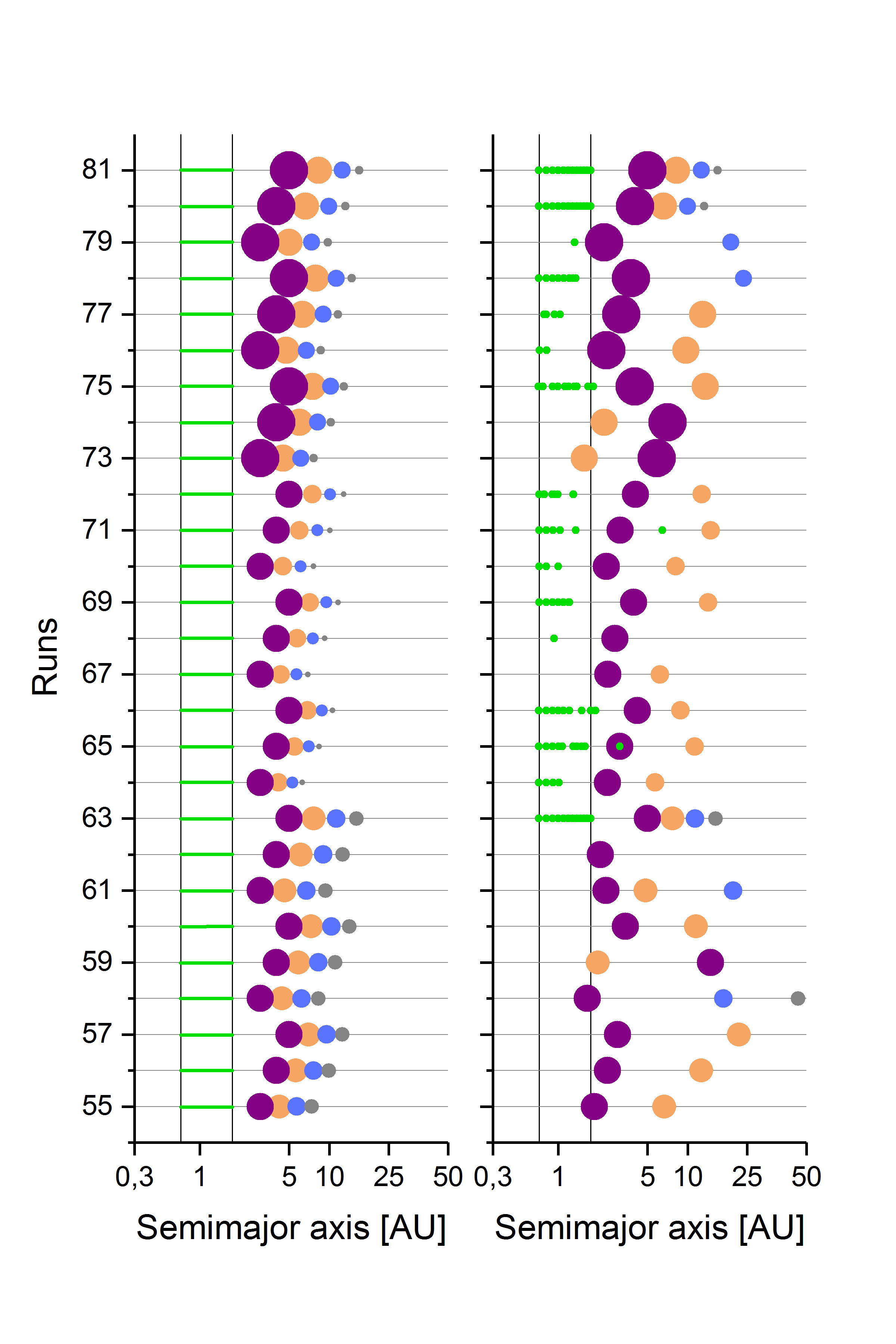}
	\caption{Same as figure 1 for four planets (runs 55-81).}
	\label{Results4pl}
\end{figure}

In figure 4 only seven runs left more than one test particle in the habitable zone. In many cases the first or the second giant migrates close to the outer border of the habitable zone. The fifth giant got always ejected except for three runs (88, 89 and 90).\\

\begin{figure}[!htbp]
	\centering
	\includegraphics[width=0.5\textwidth]{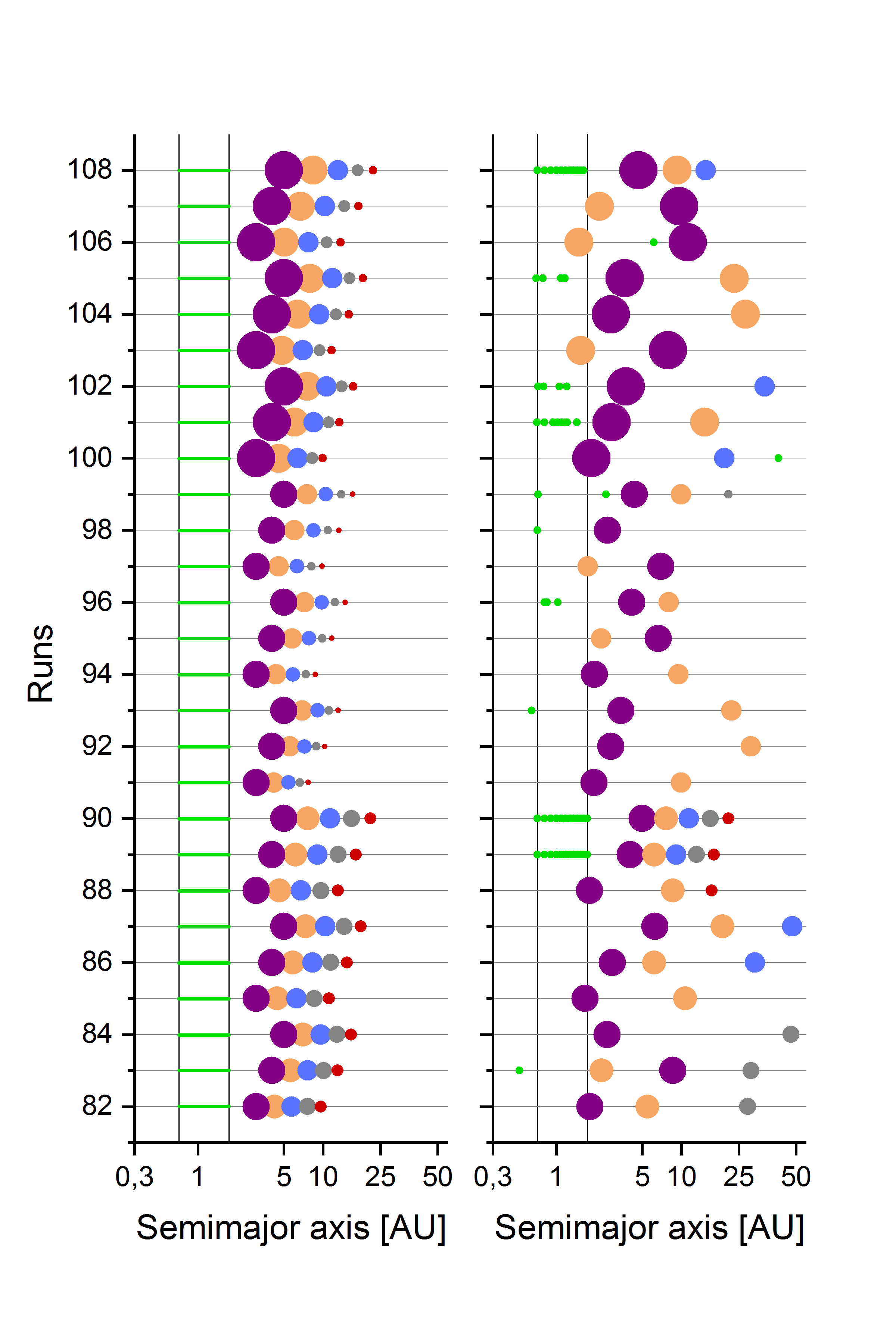}
	\caption{Same as figure 1 for five planets (runs 82-108).}
	\label{Results5pl}
\end{figure}

The simulations with initial six gas giants left over almost no habitable test particles. The outer three giants usually got ejected and the innermost planet migrated further inward (Figure 5).\\

\begin{figure}[!htbp]
	\centering
	\includegraphics[width=0.5\textwidth]{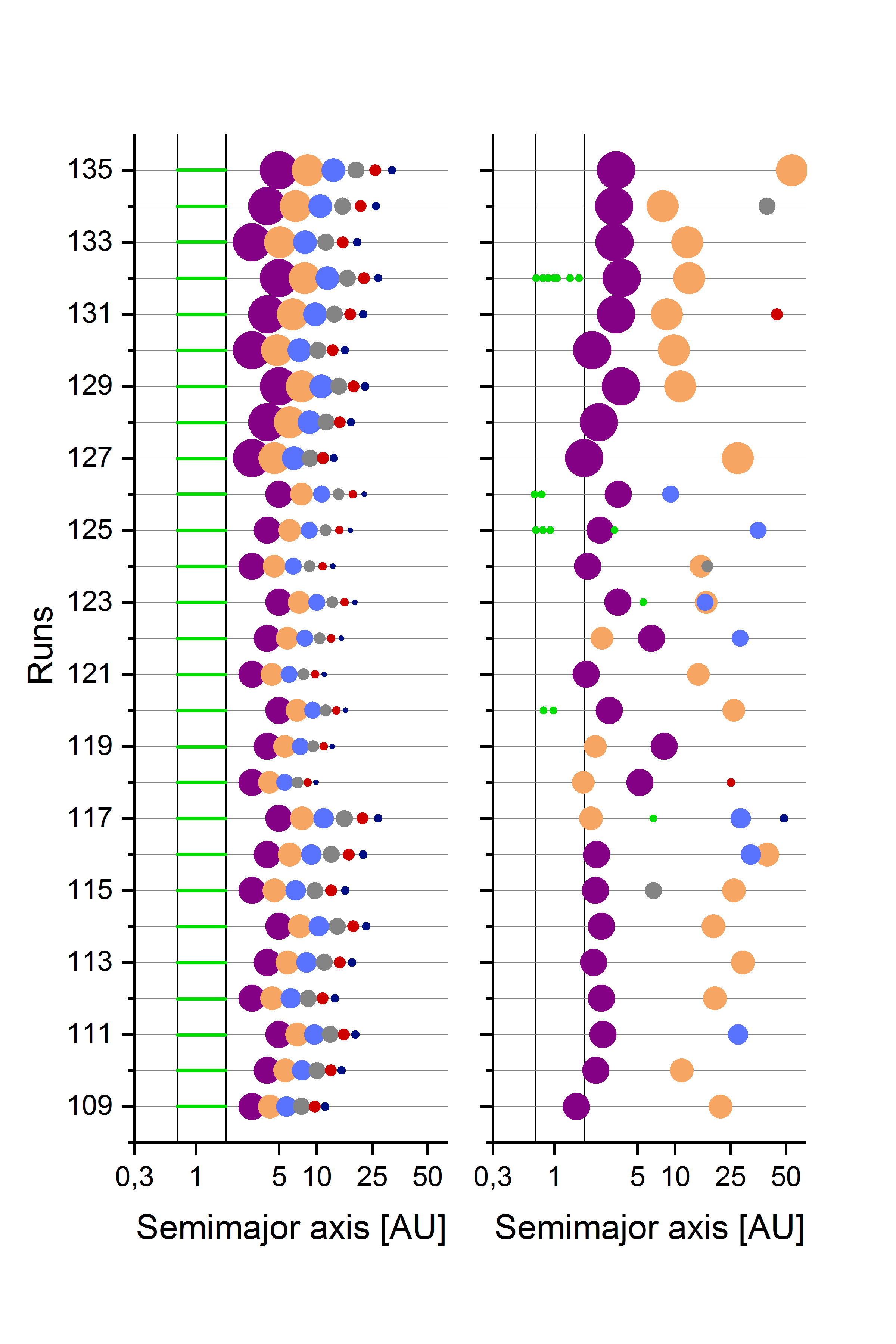}
	\caption{Same as figure 1 for six planets (runs 109-135).}
	\label{Results6pl}
\end{figure}

\paragraph{Initial Parameter Dependencies}

\begin{figure}[!htb]
	\centering
	\includegraphics[width=0.5\textwidth]{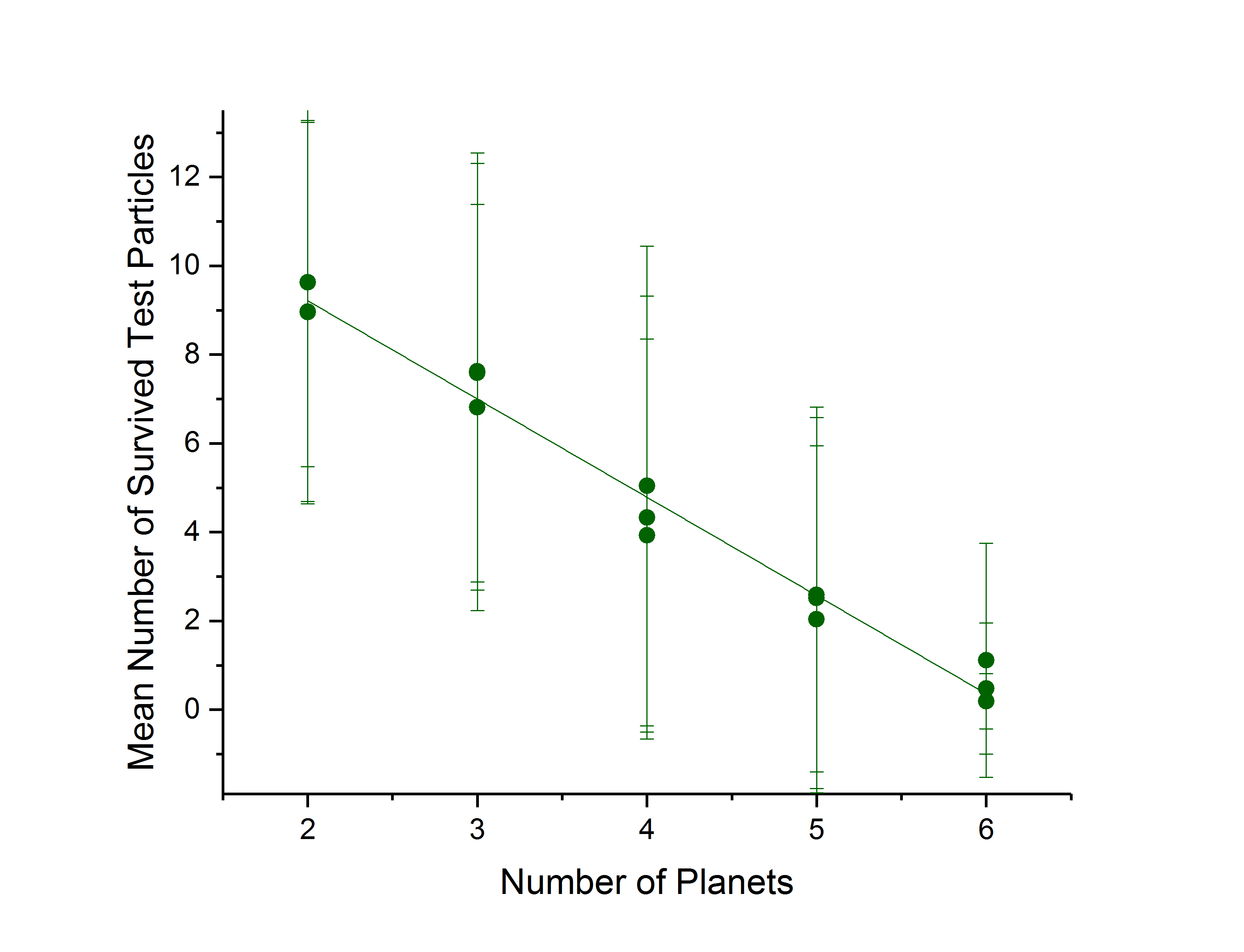}
	\caption{Mean number of surviving test particles in the habitable zone at the end of the simulations ($TP_{Hab}$) as a function of the initial number of gas giants ($N_0$). Three mean datapoints are presented for each planet number.}
	\label{TP_survivedversusN_planets}
\end{figure}

Figure 6 shows the dependancy of the mean number of surviving test particles on the initial number of giant planets. The higher the number of gas giants the lower is the number of habitable test particles. For systems with six planets almost all test particles were ejected in every simulation (on average 90 percent), which results in smaller error bars. Simulations with five initial giant planets have a survival rate of 25 percent and four initial giant planets result in a survival rate of about 40 percent. On the other hand in planetary systems consisting of two or three giant planets in average more than half of the test particles (75 percent for $N=2$ and 60 percent for $N=3$) stayed in the habitable zone. Figure 6 indicates that a linear correlation exists between the number of habitable test particles and the number of giant planets.\\

\begin{figure}[!htb]
	\centering
	\includegraphics[width=0.5\textwidth]{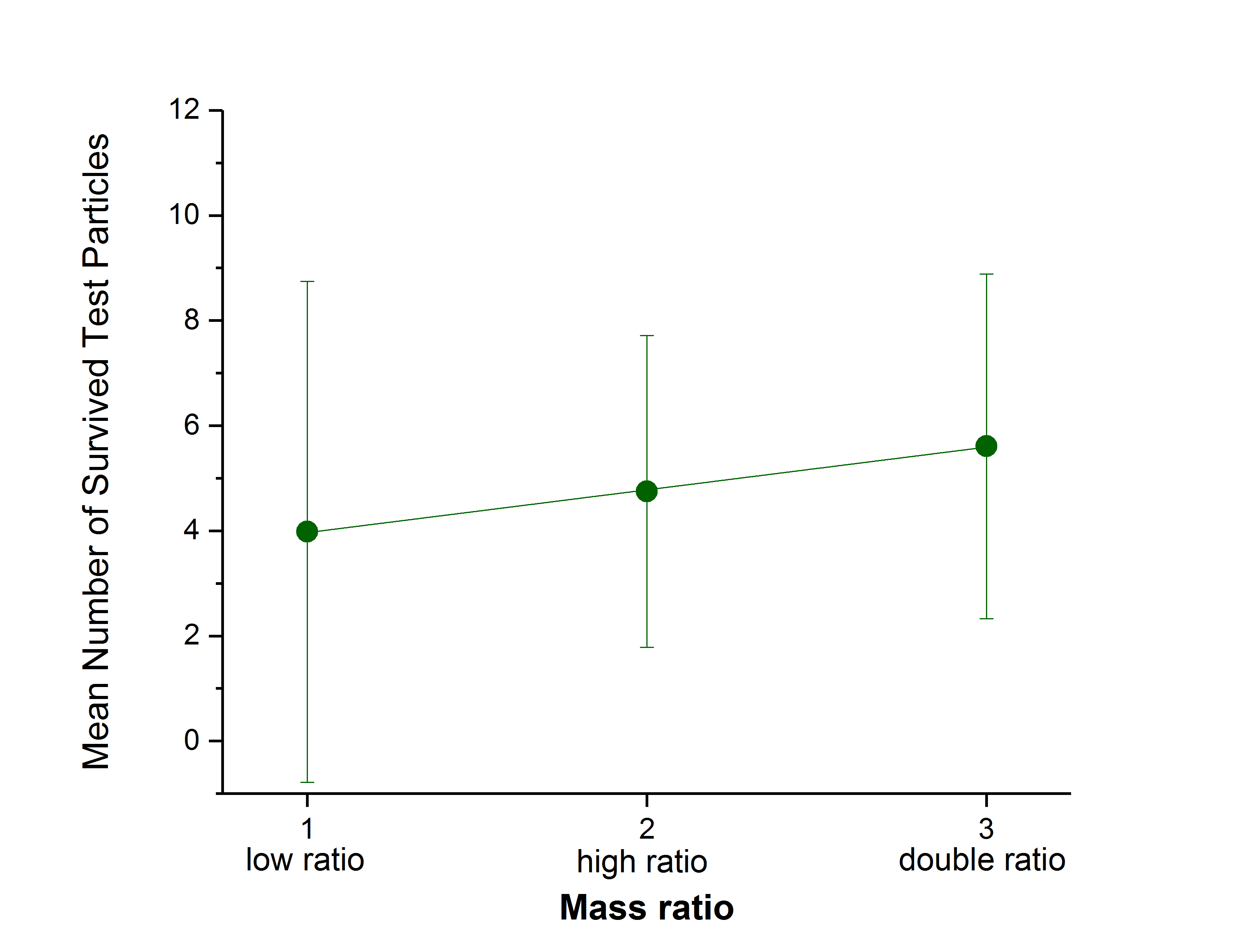}
	\caption{Mean number of surviving test particles ($TP_{Hab}$) dependent on the mass constellation (1=low ratio, 2=high ratio, 3=double high ratio).}
	\label{TP_survivedversusMassratio}
\end{figure}

Figure 7 shows a linear correlation similar to figure 6 but less steep. However, high standard deviations exist in this plot because the mean values of all runs are shown including all initial parameter variations. For a low ratio of planet masses most test particles were ejected. Similar to Carrera et al., figure 7 indicates that the survival rate of terrestrial planets increases as the giant planet masses become more hierarchical. Unexpectedly, the runs with the highest planet masses were the runs with the highest number of surviving test particles even higher than the high ratio runs. In contrast, Carrera et al. found that runs with double mass of their four planet set have the same outcome than the original runs.\\

\begin{figure}[!htb]
	\centering
	\includegraphics[width=0.5\textwidth]{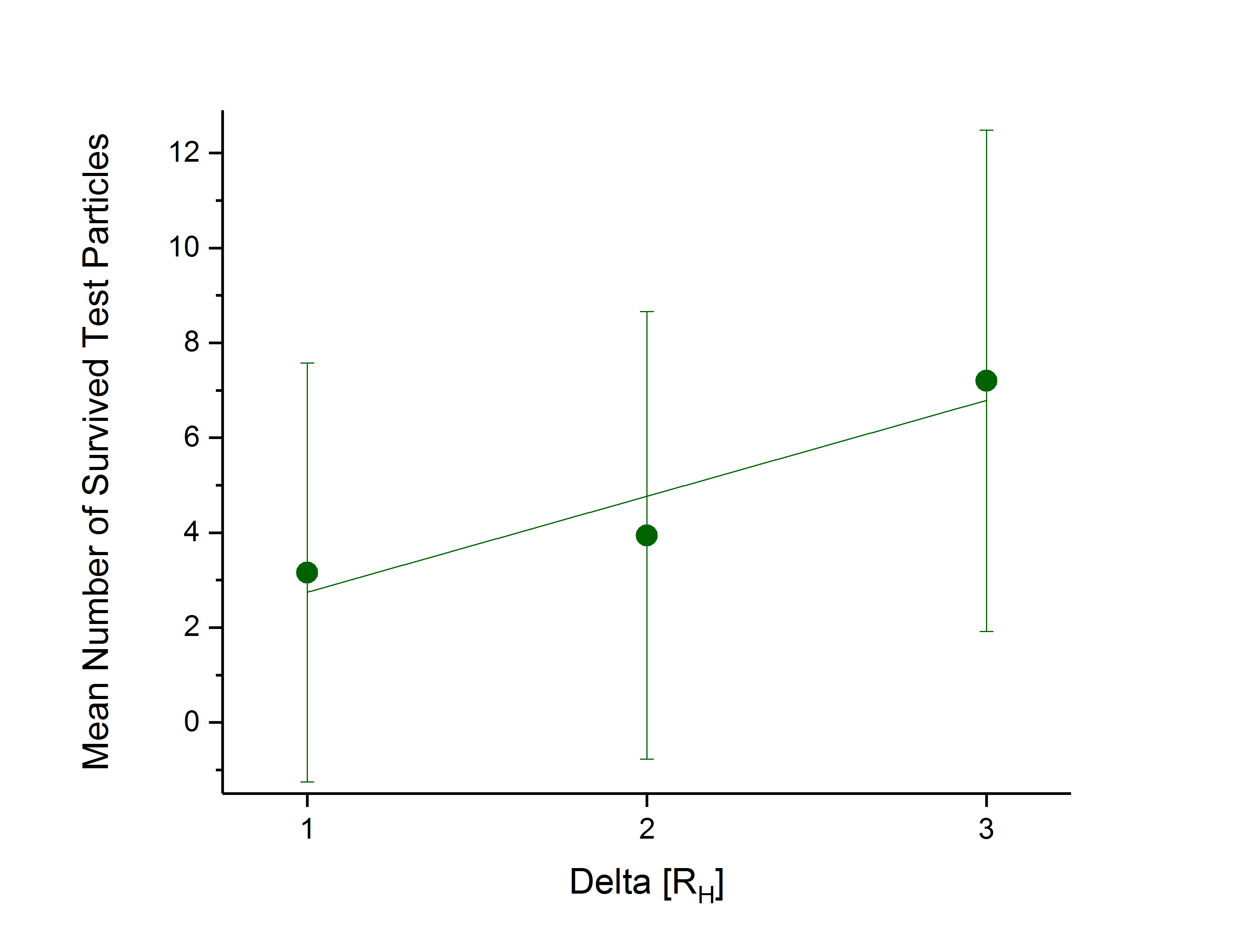}
	\caption{Mean Number of surviving test particles as a function of the distances between the giant planets in Hill radii. 1, 2 and 3 represent the three values for $\Delta$ corresponding to the planet number (Table 1).}
	\label{TP_survivedversusDelta}
\end{figure}

Figure 8 demonstrates the correlation between the surviving test particles and the distance between the giant planets expressed in Hill radii ($\Delta [R_H]$). As anticipated the more compact systems are more destructive to test particles. The middle value for the distance of 4.5 Hill radii is more destructive than expected. Systems with initial distances of 5 Hill Radii ($3.4$ for two giant planets) have on average a $60$ percent chance to have a habitable terrestrial planet, whereas systems with distances of 4 and $4.5$ Hill Radii (3 and $3.2$ for two giant planets) have a chance of $26$ percent and $30$ percent to harbour a habitable planet. \\

\begin{figure}[!htb]
	\centering
	\includegraphics[width=0.5\textwidth]{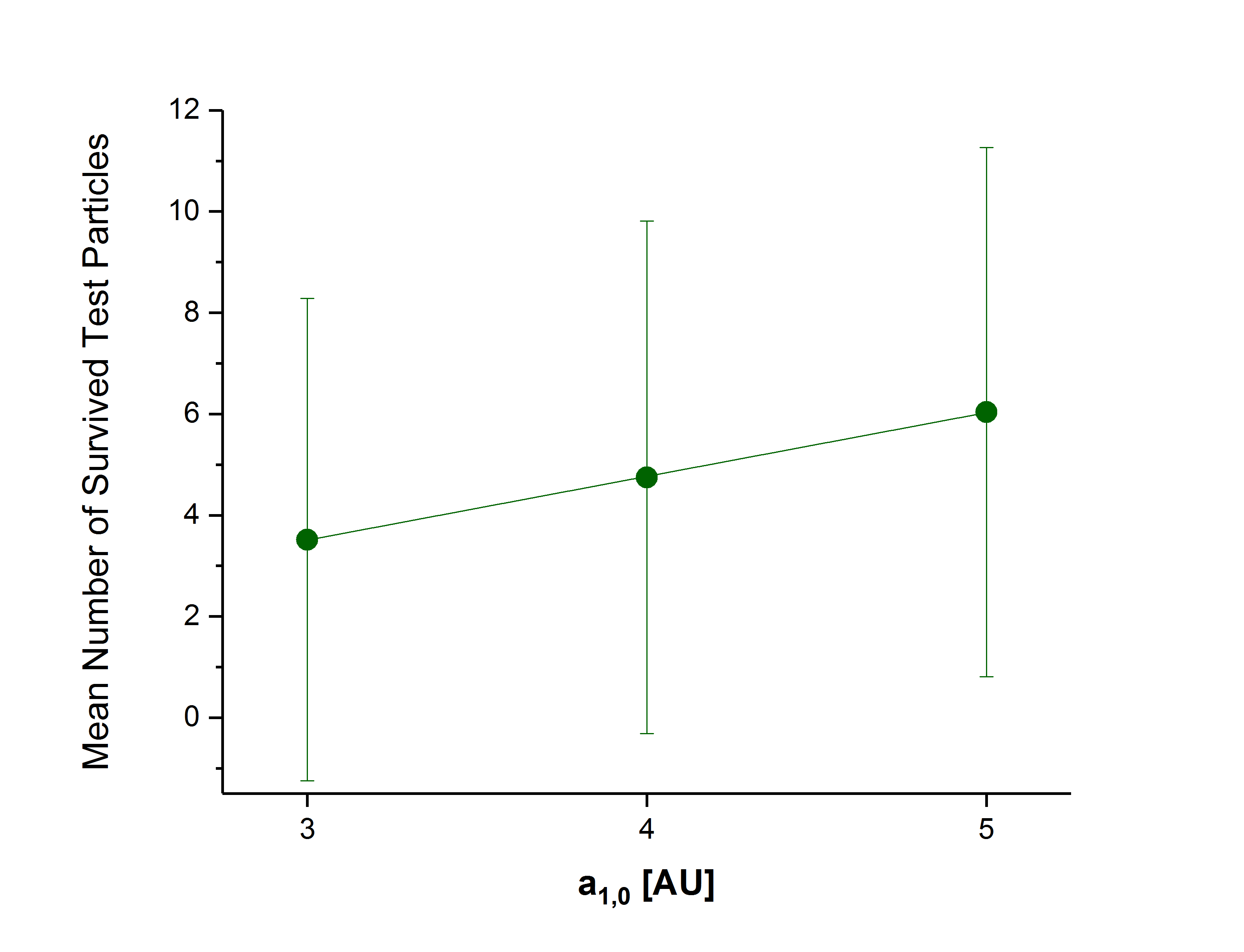}
	\caption{Mean number of surviving test particles as a function of the semimajor axis of the first giant planet $a_{1,0}$.}
	\label{TP_survivedversusa_1_0}
\end{figure}

Figure 9 shows the dependancy on the semimajor axis of the first planet in astronomical units. The results show a linear correlation with a relative low slope. Although in this plot high standard deviations are shown similar to figure 7 and 8, there exists a clear trend to higher survival rates with larger distances of the first gas giant. The runs with the nearest orbit of the first planet at 3 AU result in the lowest survival rate. This can be explained by the lower distance of the giant planets to the habitable zone.\\\\

\begin{figure}[!htb]
	\centering
	\includegraphics[width=0.5\textwidth]{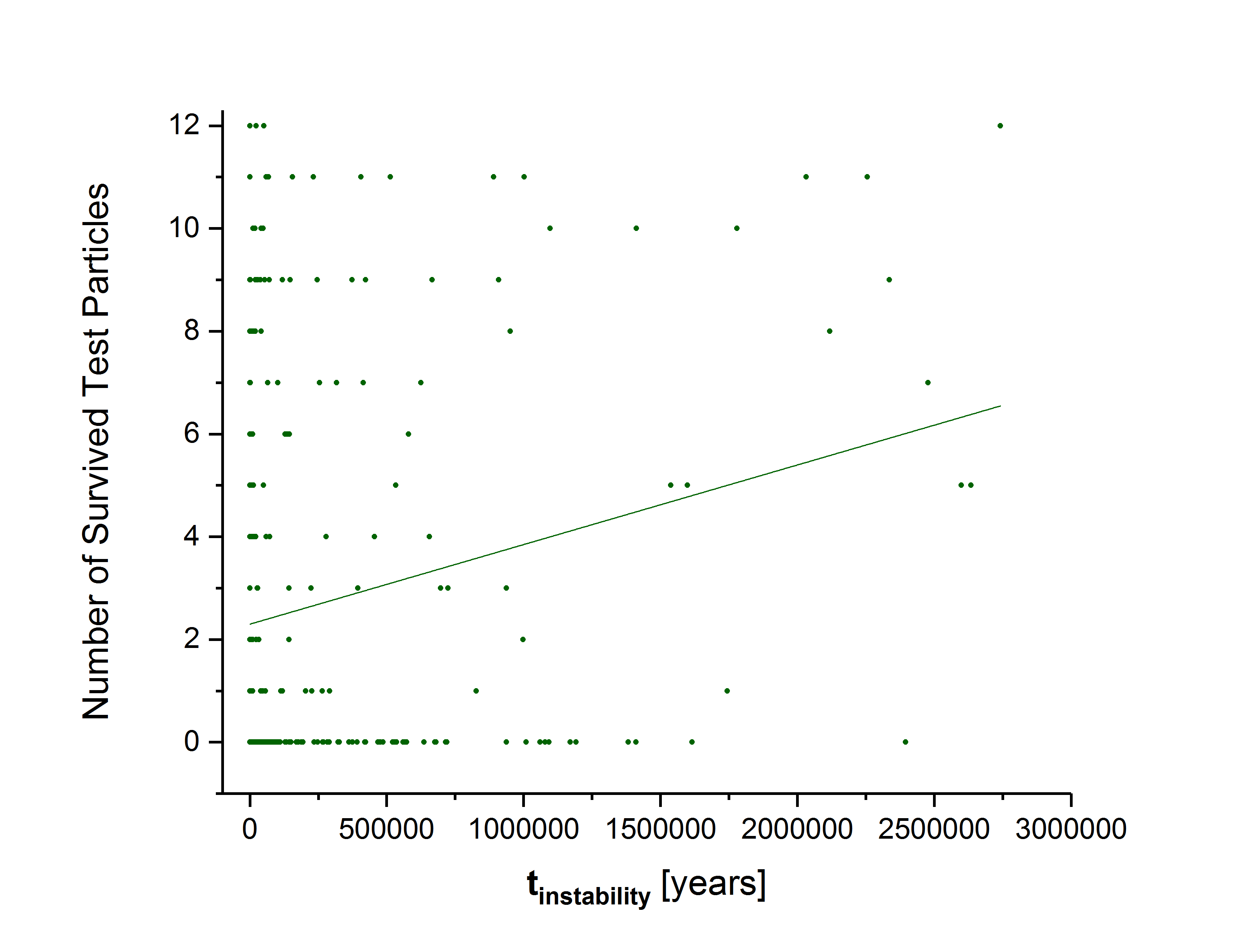}
	\caption{Mean number of surviving test particles as a function of the time at which the fist close encounter occured $t_{instability}$.}
	\label{TP_survivedversust_CE}
\end{figure}

Figure 10 shows the number of surviving test particles versus the time of the first close encounter of each run. The time of the first close encounter indicates the start of the instability phase. The linear fit shows that the trend goes to higher survival rates at later instability phases considering the relation between high and low numbers of surviving test particles for early and late instability phases. It is also seen that most instability phases took place very early in the first 250000 years of the simulations. This may be a result of too compact initial systems, which become instable very soon after starting the simulation. The more realistic runs are those with late instability phases occuring between one and three million years. These runs have the highest survival rates and lowest ejection rates.\\

\begin{figure}[!htb]
	\centering
	\includegraphics[width=0.5\textwidth]{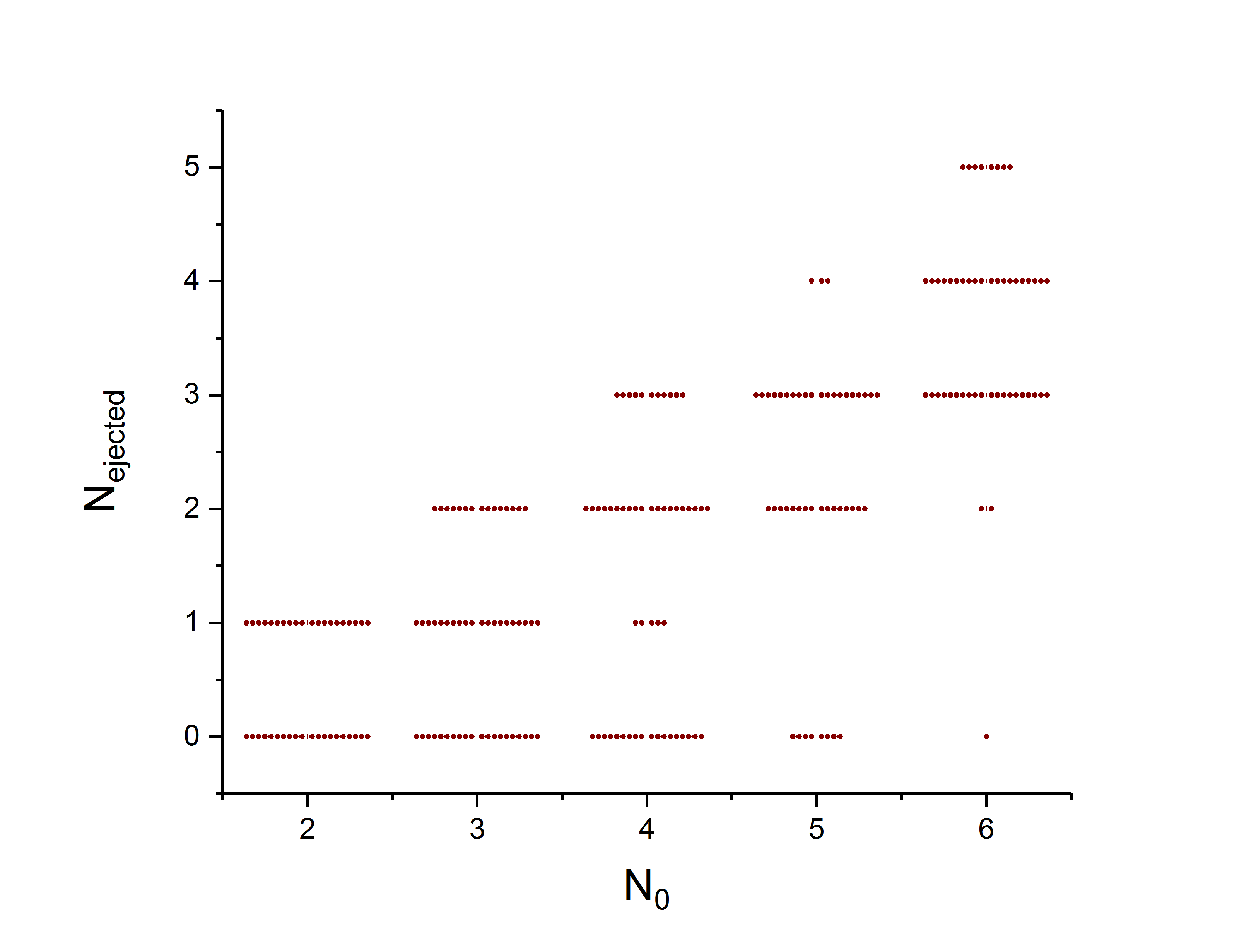}
	\caption{Ejected giant planets $N_{ejected}$ as a function of the initial number of giants $N_0$.}
	\label{N_ejectedversusN_0}
\end{figure}

Figure 11 shows how many giant planets were ejected during the simulations. Each data point represents one simulation with two, three, four, five or six initial planets. In systems with five or six initial giant planets in almost every run more than two planets were ejected. In three planet systems about one third of all runs eject two planets, one third eject one planet and the last third remains complete. Interestingly, it is very unlikely that in four, five and six giant systems one planet was ejected. Especially in five and six giant systems that scenario happened not a single time. In two planet systems in about half of the runs the second planet was ejected except for one run, where the first planet got ejected. In total the first planet got ejected only in five simulations. Four of those are runs with six initial planets.\\

\paragraph{Different Star Types}

To analyse whether the stellar mass has an impact on the stability of orbits within the habitable zone, additional runs with different stellar masses have been simulated. Therefore, runs 37 to 45 were chosen, which are runs with initially three giant planets with a high mass ratio. These runs were simulated two times again with $0.8$ $M_{sun}$ (406-414) and $0.3$ $M_{sun}$ (415-423). This set of runs has a relatively high survival rate of test particles.\\

In figure 12 the results of this additional set with different stellar masses is shown. At first sight, runs 406 to 423 have very high survival rates but also high variations. In almost all cases at least one giant planet was ejected. Compared to the runs 37 to 45 with one solar mass, there are two exceptional cases where almost all test particles were ejected (412 and 422). Figure 12 shows that in nearly all simulations with $0.8$ and $0.3$ solar masses instability phases occured (exceptions are 413 and 414). For $0.3$ solar masses the test particles typically move a little outward to the outer border of the habitable zone. Interestingly, the runs with $\Delta=5$ are the most stable cases for one solar mass, but for smaller solar masses these cases have the lowest survival rates.\\ 

\begin{figure}[!htb]
	\centering
	\includegraphics[width=0.5\textwidth]{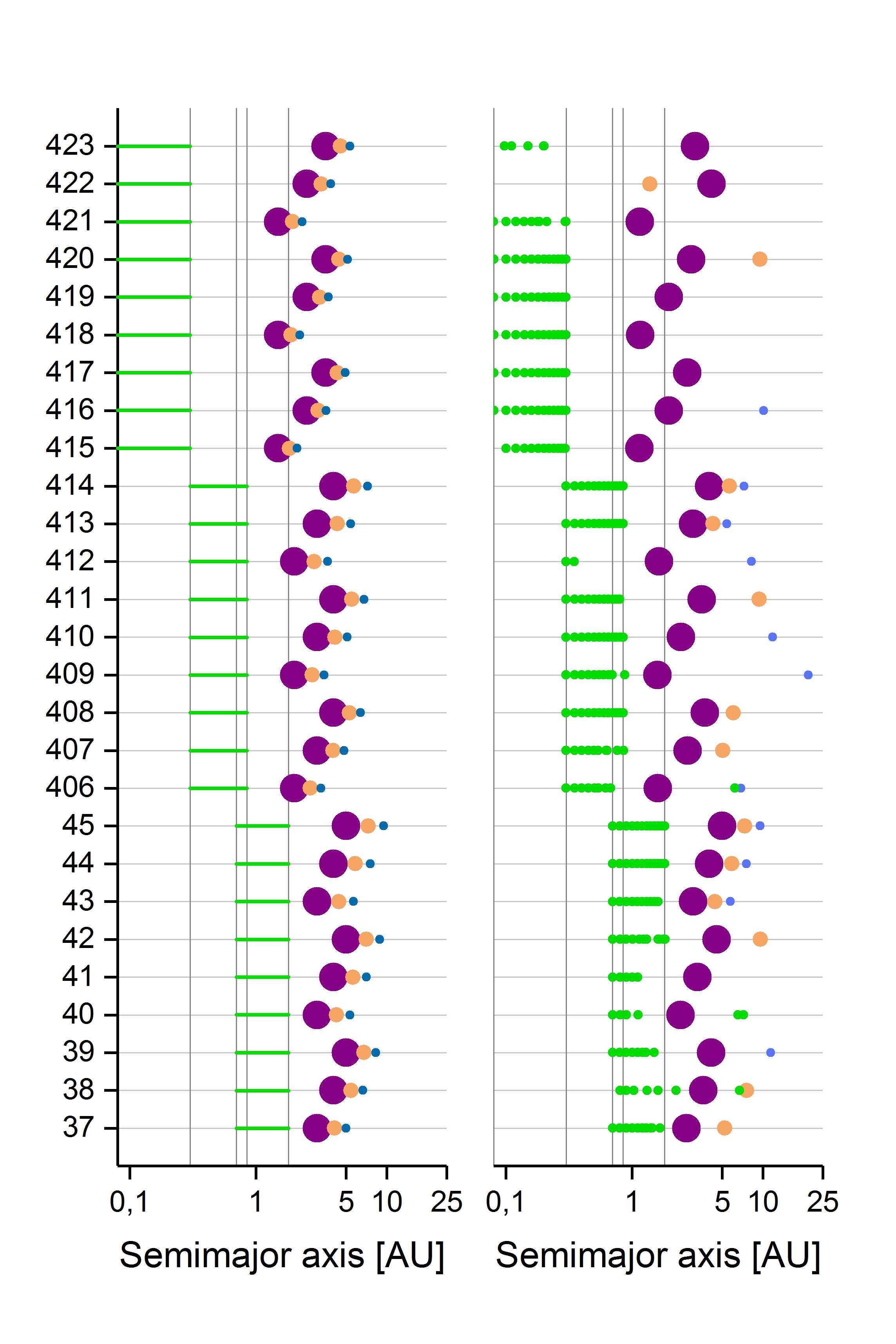}
	\caption{Illustration of the runs 37 to 45 and 406 to 423. The left panel shows the initial state and the right panel shows the final state. The position in astronomical units is represented through a logarithmic x-axis and the size of the circles is proportional to the planet mass. The vertical grey lines indicate the borders of the habitable zones (Table 3), respectivly. The green lines (left panel) and the green dots (right panel) represent the twelve test particles.}
	\label{Vorher_Nachher_Mass_star}
\end{figure}

Figure 13 shows the mean number of surviving test particles dependent on the stellar mass. The plot reveals no linear correlation between the three different stellar masses. $0.3$ and $0.8$ solar masses show higher mean survival rates ($9.56$ and $9.89$) than one solar mass ($7.78$), whereas for $0.3$ solar mass the highest variations exist.\\

\begin{figure}[!htb]
	\centering
	\includegraphics[width=0.5\textwidth]{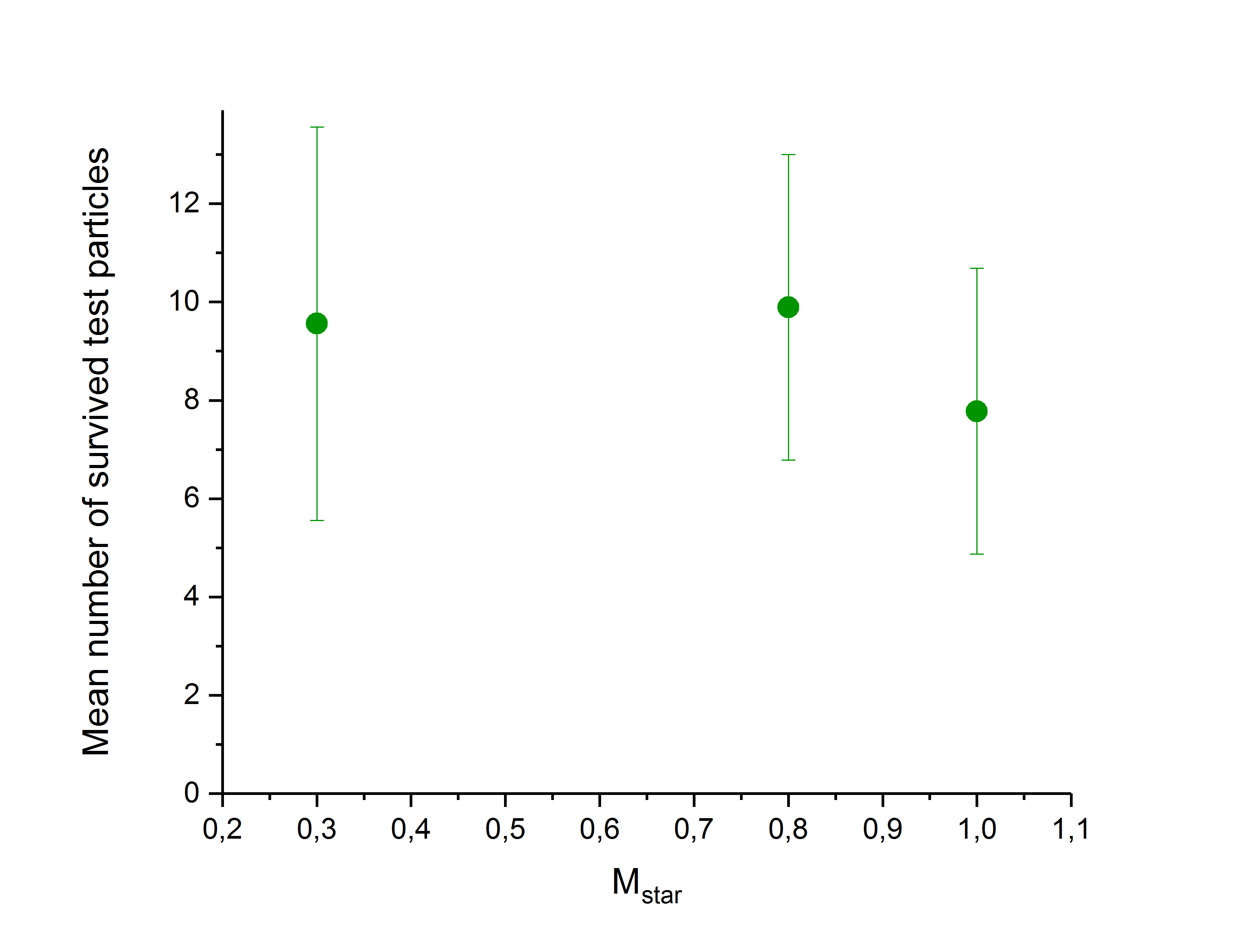}
	\caption{Mean number of surviving test particles as a function of the stellar mass $M_{star}$}
	\label{N_habversusMass_star}
\end{figure}

\paragraph{End Parameter Dependencies}

In this section the correlation of the survival rate of test particles and the final orbital parameters (semimajor axis, eccentricity and inclination) of the giant planets will be explored. The final parameters of the simulations represent the present-day orbits of detected gas giants.\\

In figure 14 (Appendix) each layer illustrates the final eccentricity of the respective giant planet as a function of the final semimajor axis. The survival rate of the test particles is represented through the colours of the dots. The first layer on the top left side shows that in most runs the first giant planet migrates inward, which leads to a very low survival rate (red data points). Outward migration of the first giant further than 5 AU results in orbit crossings and interactions with the other planets and lead also to a very low survival rate. For semimajor axes between $2.5$ and 5 AU and eccentricities lower than $0.35$, there is a good chance for the test particles to survive.\\

For the second planet (layer on the top right side) inward migration interior to 4 AU and outward migration exterior to 17 AU results in no surviving test particles. For eccentricities lower than $0.1$ and semimajor axes between 4 and $8.5$ AU almost all test particles survived. The third giant planet (middle left) shows the highest survival rates for semimajor axes between 5 and 15 AU and eccentricities lower than $0.1$. Same counts for the fourth planet (middle right), but for semimajor axes between 9 and 17 AU. In the two bottom layers not many data points exist, because in most of the runs with five or six initial planets the fifth and the sixth giant got ejected or migrated outward.\\

Overall migrations over large distances usually led to low survival rate of test particles. As a point of reference, giant planets with eccentricities larger than $0.4$ probably have no habitable terrestrial planets, which is in line with the conclusion of Carrera et al.\\

Figure 15 (Appendix) shows the final inclination of each giant planet as a function of the respective semimajor axis. The two layers on the top corresponding to the first and second giant planet show that these planets could reach very high inclinations especially as a result of inward migration. Test particles could survive for inclinations lower than 20 degrees. For the third and fourth planet exist exceptions with inclinations higher than 20 degree and surviving test particles. \\

\section{\label{sec:model} Discussion and Conclusions}

423 simulations have been made to explore the fate of habitable terrestrial planets in planetary systems with several giant planets. Those systems often experience dynamical instabilities within the simulated three million years, which matches with the observed high eccentricities of giant exoplanets. Such instabilities can alter the orbital parameters of other planets in the system and may also take a terrestrial planet out of the habitable zone. In the simulations terrestrial planets were treated like test particles. Most of the graphs show the dependency of the mean number of habitable test particles on different initial and final parameters of the system.\\

A linear correlation between the number of giant planets and the number of habitable test particles was found. Planetary systems consisting of two giant planets were fairly benign to terrestrial planets, whereas six giant planets very often lead to a complete clearing of the habitable zone. Similar to Carrera et al. the results show that the more hierarchical systems regarding the planet mass lead to a higher survival rate of test particles. However, doubling of the planet masses results in even higher survival rates of terrestrial planets, contrary to Carrera et al., who had no different outcome with double mass giant planets. As anticipated the dependency on the distance between the giant planets expressed in Hill radii is almost linear. The larger the distance the higher is the stability of the system and the survival rate of test particles. However, the middle value of $4.5$ Hill Radii leads to a little lower survival rate than expected. Analogous, the dependency on the semimajor axis of the first giant planet seems to be linear. The larger the initial orbits of the giant planets are the higher is the survival rate of terrestrial planets due to the larger distance to the habitable zone.\\ 

Furthermore, it was found that the survival rate increases with the time of the instability phase. Instabilities occuring in the first 100000 years after the beginning of the simulation are very destructive to the habitable zone. Those planetary systems are mostly very compact initial systems, which may be unrealistic. Instability timescales between one and three million years were only achieved with a $\Delta$ of five Hill radii, which corresponds to the largest initial distance between the planets. Those runs might be more realistic and led usually to a higher number of surviving habitable planets. In continuative research one could run simulations with wider initial systems for longer than three million years. The ejection of one or more giant planets is a common scenario. Especially, planetary systems consisting of five or six giant planets are very chaotic and destructive. However, this number of initial giant planets resulted in none of the simulations in the ejection of exact one giant planet.\\

The additional simulations with $0.3$ and $0.8$ solar masses (M- and K-stars) show that those systems have higher mean survival rates about $80$ percent and $83$ percent compared to systems with one solar mass, which have a mean survival rate of $65$ percent for habitable planets. Overall the number of planets and the distance between these as well as the distance of the giant planets to the habitable zone play a major role in the dynamical evolution of habitable terrestrial planets. The planet masses and the stellar mass have a smaller impact on the stability of habitable planets.\\

Similar to Carrera et al. a link between the present-day orbit of an observed giant exoplanet and the survival of habitable planets was established. The final parameters of the simulations represent the present-day orbital parameters of observed giant exoplanets and the mean number of surviving test particles represent the probability that a terrestrial planet exists in the habitable zone. It is very hard to define a probability based on the observed semimajor axis of an exoplanet, because one does not know the initial position of the giant exoplanet after the planet formation process. In the simulations final semimajor axes of gas giants lower than $2.5$ astronomical units usually led to a low survival rate of test particles. Overall, migration over large distances, which corresponds to stronger planetary interactions and instabilities, leads to a reduction of the survival rate of habitable planets.\\

In contrary to the final semimajor axis, the final eccentricity and inclination of gas giants gives a clearer probability of habitable planets, because planetary systems usually end up with very low eccentricities and almost coplanar orbits after the planet formation process. The results reveal that observed giant planets with eccentricities higher than $0.4$ and inclinations higher than 20 degrees have experienced strong planet-planet scatterings and are unlikely to have a habitable companion in its system.\\

In further research one could do more runs per a specific initial condition for statistical purpose. Unfortunately, in this work the runs per initial condition were limited to three runs due to long computing time. Future observations of protoplanetary disks and simulations of planet formation will hopefully give a deeper insight into the final configuration of planetary systems after the formation process. This could limit the initial conditions for such simulations and gives a better estimate of how many habitable planets like the Earth exist in the universe. The detection of terrestrial planets in the habitable zones of exoplanetary systems with gas giants can verify the outcomes of this work and will hopefully be reached with future space missions. This can answer the question whether the solar system is common in the universe or an exceptional case of a planetary system.\\

\section{Acknowledgements}

Firstly, I would like to express my sincere gratitude to my advisor J\"urgen Schaffner-Bielich, who made this work possible and always supported me during my master study with helpful advice. My sincere thanks also goes to my second advisor Stefan Schramm. Besides my advisors, I would like to thank Christopher Czaban, who helped me with the simulations with his expert programming knowledge. Computer simulations were performed using resources provided by the Center for Scientific Computing (CSC) at the Goethe University in Frankfurt.

\makeatletter
\renewcommand*\@biblabel[1]{}

\def\bibsection{\section*{\refname}}

\section{Appendix}

\begin{figure*}[!b]
	\centering
	\includegraphics[width=0.8\textwidth]{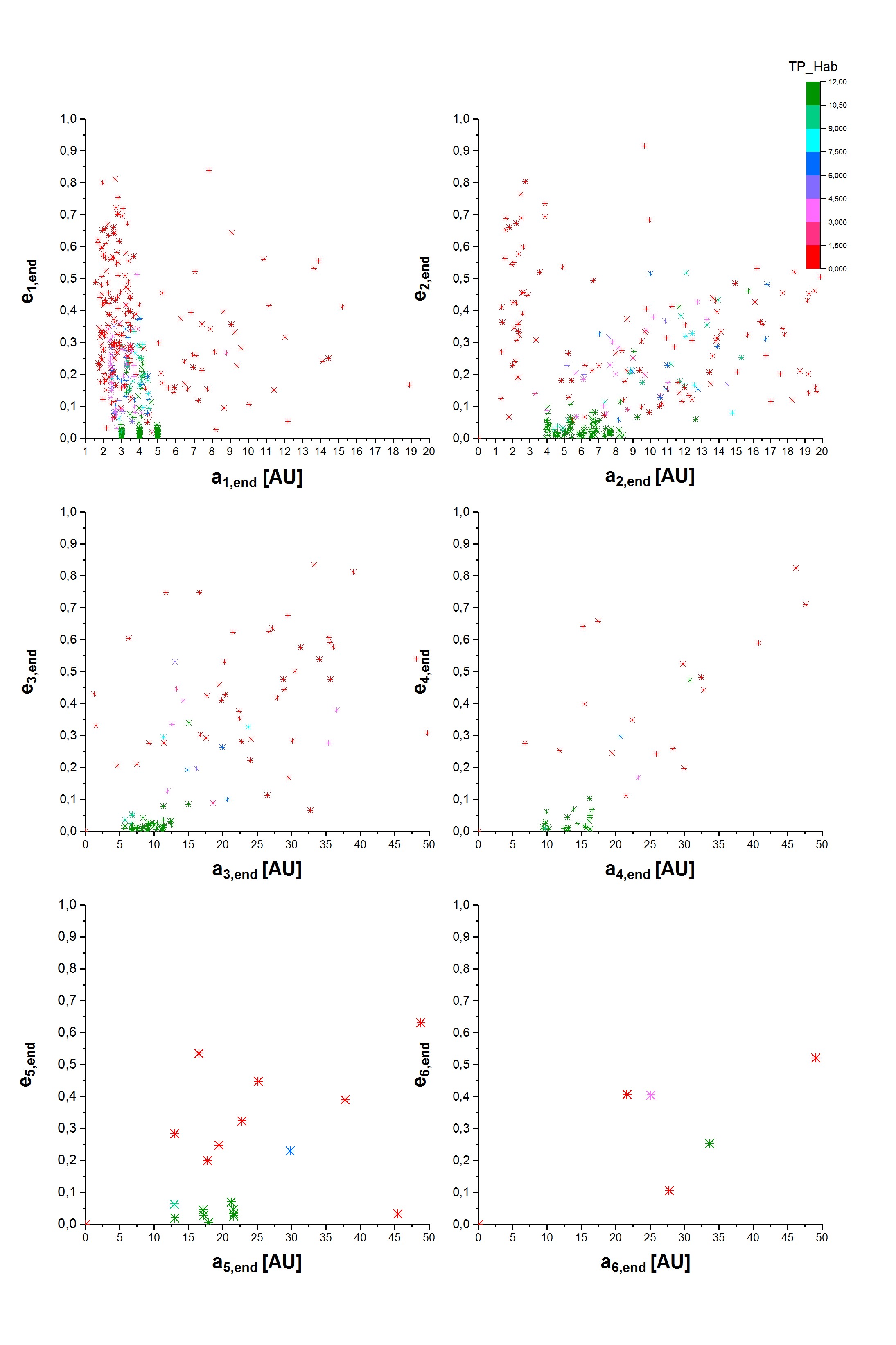}
	\caption{Final eccentricity $e_{end}$ versus final semimajor axis $a_{end}$ of each gas giant. The surviving test particles are represented through the colours of the dots.}
	\label{TP_hab_endparameter}
\end{figure*}

\begin{figure*}[!htpp]
	\centering
	\includegraphics[width=0.8\textwidth]{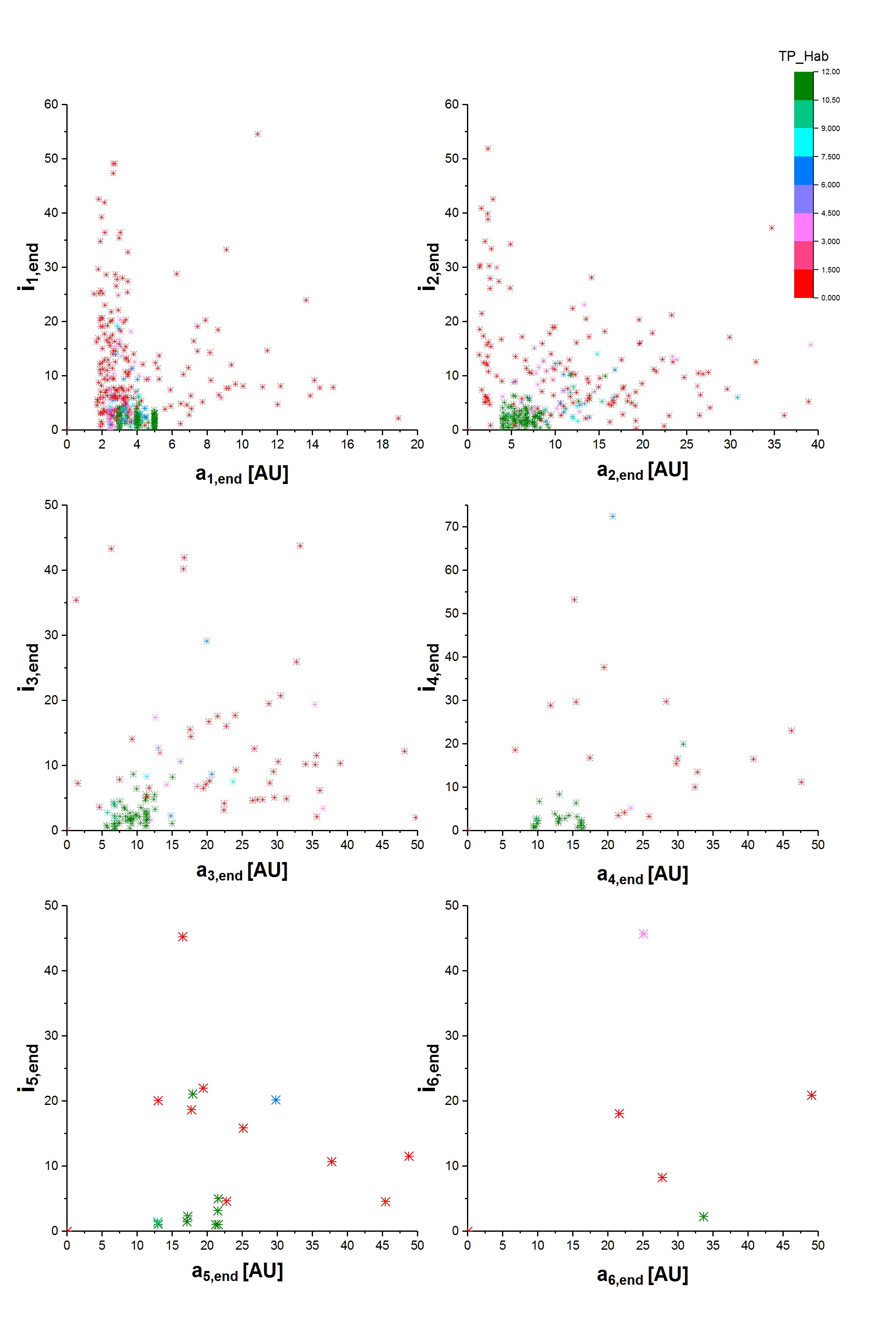}
	\caption{Final inclination $i_{end}$ versus final semimajor axis $a_{end}$ of each gas giant. The surviving test particles are represented through the colours of the dots.}
	\label{TP_hab_endparameter_inclination}
\end{figure*}

\end{document}